\documentstyle[12pt]{article}
\input epsf
\begin{document}
\begin{titlepage}
\hfill TUW--95--09 \\
\hfill gr-qc/9511009 \\

\vfill
\begin{center}
{\LARGE Geometric Interpretation\\
and Classification of Global Solutions\\
\vspace{3.5mm}
in Generalized Dilaton Gravity}\\[0.7cm]
\vfill
\renewcommand{\baselinestretch}{1}
{\large
{M.O.\ Katanaev\footnote{Permanent address:~Steklov Mathematical Institute,
Vavilov
St., 42, 117966 Moscow, Russia,  \\ katanaev@class.mian.su}}}\\
Erwin Schr\"odinger International Institute \\
for Mathematical Physics\\
Pasteurgasse 6/7, A-1090 Wien\\
Austria\\[0.7cm]
{\large
{W.\ Kummer\footnote{wkummer@tph.tuwien.ac.at} and H.\
{Liebl\footnote{liebl@tph.tuwien.ac.at}}}}\\
Institut f\"ur Theoretische Physik \\
Technische Universit\"at Wien\\
Wiedener Hauptstr. 8-10, A-1040 Wien\\
Austria\\
\end{center}
\vfill

\begin{abstract}
Two dimensional gravity with torsion is proved to be
equivalent to special types of generalized 2d dilaton gravity.
E.g. in one
version, the dilaton field is shown to be expressible by the extra scalar
curvature, constructed for an independent Lorentz connection corresponding
to a nontrivial torsion. Elimination of that dilaton field yields an
equivalent torsionless theory, nonpolynomial in curvature. These theories,
although locally equivalent exhibit
quite different global properties of the general solution.
We discuss the example of a (torsionless) dilaton theory equivalent
to the $R^2 + T^2$--model.
Each global solution of this model is shown to split into a set of
global solutions of generalized dilaton gravity.
In contrast to the theory with torsion the
equivalent dilaton one exhibits
solutions which are asymptotically flat
in special ranges of the parameters.
In the simplest case of ordinary dilaton gravity we clarify the well kown
problem of removing the Schwarzschild singularity by a field redefinition.
\end{abstract}
\vfill

Vienna, October 1995 \hfill
\end{titlepage}
\vfil

\newpage

\section{Introduction}
The renewed interest in two dimensional gravity models can be
traced to the advent of string theories \cite{wit91} with the
especially promising aspect to study dynamical models of black
holes.  Generalized dilaton theories \cite{ban91} or equivalent
theories with higher powers in curvature with the dilaton
eliminated, have enriched our knowledge  about even more complicated
singularity structures.
However, the common aspect of the latter generalizations, as well
as of other, simpler models of 2d--gravity \cite{bar79}, is the
fact that in most of these cases a physical or geometrical
interpretation of particular models does not seem to be evident.  On
the other hand,  a corresponding interpretation obviously is present
in theories
quadratic in curvature $R$ and in torsion $T$ \cite{kat86}: Such
a theory resembles a (non compact) gauge theory with second order
field equations in the zweibein $e_\mu^a$ and the Lorentz connection
${\omega_\mu^a}_b$, if the latter is considered as an independent
variable.  The globally complete solutions of those by now well
studied models with torsion \cite{kat86,sch93} show a
rich singularity structure \cite{kat93}.

A common feature of all such models, though, --- in the matterless
case --- is the large number of unphysical degrees of freedom.
Therefore, the question arises whether the (classical) equivalence
between generalized dilaton theories and higher derivative
curvature models \cite{ban91} in the torsion free case may not
even extend to theories with nonvanishing torsion, thereby
yielding a geometric interpretation of certain dilaton theories
\cite{kaw93}.

This seems to be especially attractive in view of the fact that a
singularity resembling most closely the black hole in $d = 4$ has
been the main motivation for the interest in dilaton theories. On
the other hand, after coupling matter fields to the original
dilaton black hole, so far the hopes did not materialize that e.g.
the problem of the depletion of a black hole by Hawking radiation
can be understood in more than a semiclassical way, nor the even
more fundamental question of information loss \cite{kaw93}.
Thus, by extending the range of 2d--models which are classically
solvable and which contain the Schwarzschild singularity among
their global solutions, possibly --- after introducing couplings to
matter --- some new model may show advantages also in the quantum
case. \\

Here we only present the first step in such a program. We show how
a theory quadratic in torsion and curvature may indeed be
reformulated as an equivalent dilaton theory, first (Section 2)
by a suitable identification of the scalar curvature in
$R^2+T^2$--theory with the dilaton field of generalized dilaton
theory.  In this case the use of the field equation necessitates a
careful check of the admissibility of the (nonlocal)
transformation involved.  In Section 3 we show this equivalence
by a {\sl local} method starting from a first order formalism for the
$R^2+T^2$--theory \cite{sch93}.  Here the Lorentz connection is
eliminated in favor of the torsion which turns into a nondynamical
field variable.
After the discussion of the general solution (Section 4) we present the
classification in Section 5.
Essential steps of the corresponding mathematical analysis are given
in Section 6. The relation of the global solutions for the dilaton
theory equivalent to the $R^2+T^2$--model \cite{kat86} we present in
Section 7. In the final Section 8 -- beside a summary -- also the role of
field redefinitions and the creation of the singularity for the ordinary
dilaton theory is discussed.
\section{Nonlocal Equivalence}

The basis of our considerations  is the Lagrangian of two dimensional gravity
 with
torsion,  containing two coupling constants
$\alpha,~\beta$ and a cosmological constant $\Lambda$ ($e = \det e^a_\mu$)
\cite{kat86}

\begin {equation}                                       \label{Gl:1}
  {\cal L}_G=\frac{e}{8\beta} R^2-\frac{e}{2\alpha} T_\alpha T^\alpha+e\Lambda.
\end {equation}

The scalar curvature $R$ is  constructed from an independent
zweibein $e_\alpha{}^a$, $\alpha=0,1$, $a=0,1$ and Lorentz connection
$\omega_\alpha$
\begin {equation}                                       \label{Gl:2}
\partial_\alpha\omega_\beta - \partial_\beta\omega_\alpha =
-\frac12\epsilon_{\alpha\beta}R
\end{equation}
\begin{equation} \label{Gl:2a}
 \epsilon^{\alpha\beta}=\epsilon^{ab}e_a{}^\alpha e_b{}^\beta =
\frac{-\tilde\epsilon^{\alpha\beta}}{e}
\end{equation}
Here $\epsilon_{ab}=-\epsilon_{ba},~\epsilon_{01}=1$ is the totally
antisymmetric tensor, $\tilde\epsilon^{\alpha\beta}$ a corresponding tensor
density.  Greek indices denote world tensors
while Latin indices describe tensors transforming under local Lorentz
rotation. The zweibein simply connects Greek indices with Latin
ones and vice versa. The torsion
\begin {equation}                                       \label{Gl:3}
T^a_{\alpha\beta} = \partial_\alpha e^a_\beta - \partial_\beta e^a_\alpha +
(\omega_\alpha e^b_\beta -\omega_\beta e^b_\alpha) \epsilon^a_b
\end{equation}
is  expressed in terms of zweibein and connection. Its trace reads
\begin{equation} \label{Gl:4}
  T_\alpha=e_b{}^\beta(\partial_\beta e_\alpha{}^b
  -\partial_\alpha e_\beta{}^b)+\omega_b\epsilon^b{}_\alpha.
\end{equation}

Now we may interpret the definition of the scalar curvature
(\ref{Gl:2})
as an equation for the Lorentz connection. In two dimensions the
integrability conditions for these equations are trivially satisfied.
So the scalar curvature and metric uniquely define a Lorentz connection
up to a gradient of an arbitrary scalar field $\psi$
\begin {equation}                                       \label{Gl:5}
  \omega_\alpha=\omega_{\perp\alpha}+\partial_\alpha\psi,~~~~
  \partial_\alpha \omega^\alpha_\perp=0,
\end {equation}
where $\omega^\alpha_\perp$ is the divergence free solution of (\ref{Gl:2}) and
$\psi$ can be readily identified with a local Lorentz rotation.
This shows that ---  at least locally --- an arbitrary Lorentz
connection can always be
parametrized by the scalar curvature and a Lorentz angle.
Let us stress that this happens already at the
kinematical level without using the equations of motion.

We now proceed to an alternative  formulation of model (\ref{Gl:1}) in terms of
the metric
\begin{equation}                                        \label{Gl:6}
  g_{\alpha\beta}=e_\alpha{}^ae_\beta{}^b\eta_{ab},~~~~\eta_{ab}={\rm
diag}(+-),
\end{equation}
and scalar curvature $R$. To reformulate the model we start with the original
equations of
motion for zweibein and Lorentz connection  from (\ref{Gl:1})
\begin{eqnarray}                                        \label{Gl:7}
  &&\frac{1}{2\beta}\partial_\alpha R + \frac{1}{\alpha} T_\alpha=0,
\\                                                      \label{Gl:8}
  &&\frac{1}{\alpha}\nabla_\alpha T_\beta + g_{\alpha\beta}
  \left(\frac{R^2}{8\beta} - \frac{T_\alpha
T^\alpha}{2\alpha}-\Lambda\right)=0,
\end{eqnarray}
where $\nabla_\alpha$ denotes the covariant derivative corresponding to
nontrivial torsion.
The second term in (1) now is divided into two
pieces, parametrized by a constant $a$
\begin{equation}       \label{Gl:9}
  T_\alpha T^\alpha=(1-a)T_\alpha T^\alpha
  + aT_\alpha T^\alpha,~~~~a={\rm const}.
\end{equation}
In the first term of (\ref{Gl:9}) the equation of motion
(\ref{Gl:7}) is used to express
only one factor $T_\alpha$ as a gradient of the scalar curvature.
We then integrate by parts and use the identity
$$
  2\widetilde\nabla_\alpha T^\alpha=\widetilde R - R,
$$
where a tilde sign means that the corresponding quantities (in our case
the metrical connection entering the covariant derivative and the scalar
curvature) are computed in terms of the metric for {\sl vanishing} torsion
only. In
the second term of (\ref{Gl:9}) eq. (\ref{Gl:7}) is used twice to arrive at a
kinetic term for the scalar curvature. In the resulting Lagrangian
\begin {equation}
\label{Gl:10}
{\cal L}=\frac{1}{8\beta}(1-a)\widetilde RR
  -\frac{\alpha}{8\beta^2}ag^{\alpha\beta}\partial_\alpha R\partial_\beta R
  +\frac{1}{8\beta} aR^2 + \Lambda,
\end {equation}
we may now consider metric and scalar curvature as new
independent variables. Varying (\ref{Gl:10}) with respect to the scalar
curvature
$R$ and metric one obtains new  second order equations of motion
\begin{eqnarray}
\label{Gl:11}
  &&\frac12 (1-a)\widetilde R
  +\frac{\alpha a}{\beta} \widetilde\nabla^\alpha
  \partial_\alpha R+aR=0,
\\
\label{Gl:12}
  &&(1-a)\widetilde\nabla_\alpha\partial_\beta R
  -\frac{\alpha}{\beta}a\partial_\alpha R\partial_\beta R
  +\frac12g_{\alpha\beta}\left(\frac{\alpha}{\beta}a
  \partial^\gamma R\partial_\gamma R+ aR^2+8\beta\Lambda\right)=0.
\end{eqnarray}

Although Lagrangian (\ref{Gl:10}) was obtained
from the Lagrangian of two dimensional gravity with torsion the resulting
model is, in general, inequivalent to (\ref{Gl:1}) for two reasons.
First, the transformation of Lorentz connection to scalar curvature
due to equation (\ref{Gl:2}) is nonlocal and second,
the equations of motions were used.
Nevertheless, both models can be shown to be  equivalent for $a=-1$
according to the following arguments:

It is proved easily that any zweibein and Lorentz connection satisfying
the equations of motion of two-dimensional gravity with torsion
(\ref{Gl:7}), (\ref{Gl:8}) satisfy also equations (\ref{Gl:11}),
(\ref{Gl:12}).
Metric and scalar
curvature are uniquely determined by zweibein and Lorentz connection.
In fact, taking the trace of (\ref{Gl:8}) and comparing it with
equations (\ref{Gl:11},\ref{Gl:12}) one finds that eq. (\ref{Gl:11}) is
satisfied if and only if $a=-1$. Then eliminating the torsion from
(\ref{Gl:8})
by means of eq.(\ref{Gl:7}) one indeed gets eq.(\ref{Gl:12}).

The inverse statement that (\ref{Gl:7}) and (\ref{Gl:8}) follow
from (\ref{Gl:11}) and (\ref{Gl:12}) for $a = -1$, is more subtle. Eq.
(\ref{Gl:11}) for $a=-1$ can be rewritten in the form
$$
  \widetilde\nabla^\alpha\left(\frac{\partial_\alpha R}{2\beta}+
\frac{1}{\alpha}T_\alpha\right) = 0 .
$$
Its general solution
$$
  \frac{\partial_\alpha R}{2\beta} + \frac{T_\alpha}{\alpha} =
\epsilon_\alpha{}^\beta
  \partial_\beta \varphi
$$
depends on an arbitrary function $\varphi$. Here we note that both equations
(\ref{Gl:6}) and (\ref{Gl:2}) define zweibein and Lorentz
connection up to an arbitrary local Lorentz rotation entering a
general solution of eqs.(\ref{Gl:6}) and (\ref{Gl:2}) as two
independent arbitrary functions. So without loss of generality we
may use one of these functions, say $\psi$ in (\ref{Gl:5}) to set
$\varphi=0$. Then the Lorentz rotation
of zweibein and Lorentz connection must be performed
simultaneously,
and  eq.(\ref{Gl:7}) indeed is the consequence of (\ref{Gl:11}),
and  eq.(\ref{Gl:8}) immediately follows from (\ref{Gl:12}).\\

In order to bring (\ref{Gl:10})  into the familiar
form of dilaton gravity we parametrize the scalar
curvature by a dilaton field $\phi$ and rescale the metric $(\sigma
= \pm 1)$

\begin{equation}
\label{Gl:13}
  R=\sigma 4\beta e^{-2\phi},~~~~
  g_{\alpha\beta}\rightarrow \;g_{\alpha\beta}e^{2\phi}.
\end{equation}
This transformation is clearly  canonical  and thus
leads to the equivalent generalized dilaton gravity described by the
Lagrangian
\begin{eqnarray}
\label{Gl:14}
  {\cal L}_D^{(1)} &=& e^{-2\phi}\left[\sigma \widetilde R + \left(4\sigma +
  8 \alpha e^{-2\phi}\right)
  g^{\alpha\beta}\partial_\alpha\phi\partial_\beta\phi
  - 2\beta +\Lambda e^{4\phi}\right] \sqrt{-g}.
\end{eqnarray}
This coincides
with the standard dilaton gravity Lagrangian
\cite{wit91} if $\sigma = +1$,  $2\beta = -4 \lambda^2, \Lambda = 0, \alpha =
0$.
By leaving the sign $\sigma$ in (\ref{Gl:14})
open we are able to cover both regions $R > 0$ and $R < 0$ for a
fixed sign of $\beta$.

\section{Local Equivalence}

Instead of eliminating the (independent) Lorentz connection by the nonlocal
transformation defined by eq.
(\ref{Gl:2}), its components may also be solved {\sl algebraically} in terms of
the two
independent components of the torsion represented by the Hodge dual of
(\ref{Gl:3}).
In the definition
\begin{equation}
 \label{Gl:16}
 T^\pm = (\partial_\mu \pm \omega_\mu) e^\pm_\nu\,
 \tilde\epsilon^{\mu\nu}
\end{equation}
we introduce light cone coordinates $T^\pm = \frac{1}{\sqrt{2}}(T^0 \pm
T^1)$ in the Lorentz--indices. Note that here we just retain
$\tilde \epsilon ^{\mu\nu}$, as defined in (\ref{Gl:2a}).  Eq. (\ref{Gl:1})
is equivalent to the first order action with Lagrangian
\begin{equation}
 \label{Gl:17}
{\cal L}^{(1)} = X^+T^- + X^-T^+ +X(\tilde
\epsilon^{\mu\nu}\partial_\mu\omega_\nu) - e (\alpha X^+X^- + V)
\end{equation}
where
\begin{equation}
 \label{Gl:18}
 V = \frac{\beta}{2} X^2 - \Lambda.
\end{equation}
The equations of motion for $X$ and $X^\pm$ are
\begin{equation}
 \label{Gl:eom}
 e\beta X=\frac{R}{2},~~~~e\alpha X^\pm=T^\pm
\end{equation}
and therefore $X$ and $X^\pm$ are proportional to curvature and torsion
for $\beta \neq 0$ and $\alpha \neq 0$.
It should be noticed that the subsequent steps hold for
general $V = V(X)$, i.e. a theory quadratic in torsion but with
arbitrary powers in curvature.

Now instead of $\omega_\mu$ the $T^\pm$ in (\ref{Gl:16}) are introduced as new
variables:
\begin{equation}
\label{Gl:19}
\tilde \epsilon^{\mu\nu} \partial_\mu \omega_\nu = \tilde
\epsilon^{\mu\nu} \partial_\mu \tilde\omega_\nu + \tilde
\epsilon^{\mu\nu}\partial_\mu \left(\frac{e_\nu^-T^+ + e_\nu^+
T^-}{e}\right)
\end{equation}
The first term on the r.h.s. of (\ref{Gl:19}) is proportional to a torsionless
curvature $\tilde R$,
\begin{equation}
\label{Gl:20}
\tilde\epsilon^{\mu\nu} \partial_\mu \tilde\omega_\nu = - \frac{\tilde R
e}{2}\quad .
\end{equation}
Inserting (\ref{Gl:19}) into (\ref{Gl:17}), after shifting the derivatives in
the
second term of (\ref{Gl:19}) onto X exhibits the nondynamical nature of $T^\pm$
which may
be 'integrated out' by solving their (algebraic) equations of motion.
At this point the identification $X=R/(2\beta e)$ immediately yields
(\ref{Gl:10})
With a
definition of the dilaton field similar to (\ref{Gl:13})
\begin{equation}
\label{Gl:21}
\frac{X}{2} = e^{-2\phi},~~~~X>0
\end{equation}
and after reexpressing the factors $e_\nu^\pm / e$ from the square bracket
of (\ref{Gl:19}) in terms of the inverse zweibeins  combining them into
$g^{\alpha\beta} = e^{+\alpha}e^{-\beta} + e^{+\beta}e^{-\alpha}$, the
Lagrangian ${\cal L}^{(1)}$ is found to be equivalent to
\begin{equation}
\label{Gl:22}
{\cal L}^{(2)} = \sqrt{-g} \left[- e^{-2\phi} \tilde R + 8 \alpha\cdot
e^{-4\phi}g^{\mu\nu} (\partial_\mu\phi)(\partial_\nu\phi) -
V(2e^{-2\phi}) \right]~~~.
\end{equation}
The case $X<0$ will be discussed below (cf. (\ref{Gl:invar})).
In addition, using the identity
\begin{eqnarray}
\label{Gl:23}
g_{\mu\nu} &=& e^{2\varphi}\; \hat g_{\mu\nu}\nonumber\\
\sqrt{-g} R & = & \sqrt{-\hat g}\hat R - 2\partial_\alpha \left[
\sqrt{-\hat g} \hat g^{\alpha\beta} \partial_\beta \varphi\right]
\end{eqnarray}
which also represents a local transformation, allows to write down the most
general dilaton theory equivalent to (\ref{Gl:1}):
\begin{equation}
 \label{Gl:24}
{\cal L}^{(3)} = \sqrt{-\hat g}\left[ -e^{-2\phi}\hat R + 4 \hat
g^{\alpha\beta}
(\partial_\alpha\phi)\left(e^{-2\phi}\partial_\beta\varphi +
2\alpha e^{-4\phi} \partial_\beta\phi\right)
 - e^{2\varphi}
V(2e^{-2\phi})\right]
\end{equation}
The choice $\varphi = \phi$ immediately yields the  case of $\sigma
= -1$ of (\ref{Gl:14}), while for $\varphi = -
\phi$
\begin{equation}
\label{Gl:25}
{\cal L}^{(4)} = -\sqrt{-\hat g} e^{-2\phi} \left[ \hat R +
4(1-2\alpha e^{-2\phi}) (\nabla \phi)^2 + V(2e^{-2\phi})\right]
\end{equation}
the deviation from ordinary dilaton theory ($\alpha = 0, V = 4\lambda^2$) is
most obvious. Of course, the dilaton field may be eliminated altogether as
well, if in (\ref{Gl:24}) (for constant $\alpha$)
\begin{equation}
\label{Gl:26}
\varphi = \varphi (\phi) = \alpha e^{-2\phi} = \frac{\alpha X}{2}
\end{equation}
is chosen. In that case it seems more useful to retain the variable X
instead of $\phi$:
\begin{equation}
\label{Gl:27}
{\cal L}^{(5)} = - \sqrt{-\hat g}\; \left[\frac{X \hat R}{2} + e^{\alpha X} V
(X)\right]
\end{equation}
Comparing (\ref{Gl:27}) to a torsionless theory (\ref{Gl:17}) with $\alpha = 0$
but modified
$V$, the difference now just resides in the additional exponential
$e^{\alpha X}$ \cite{XY}.

\section{General Solution of Generalized Dilaton Gravity}

The study of global properties for 2d theories is based upon the
extension of the solution which is known at first only in local
patches,  continued maximally to global ones.  The analysis uses
null--directions which become the coordinates of Penrose diagrams
which are sewed together appropriately.  The continuation across
horizons and the determination of singularities can be  based upon
extremals or geodesics.  The physical interpretation of an
extremal is the interaction of the space--time manifold with a
point like test particle, 'feeling' the metric $g_{\alpha\beta}$
through the Christoffel symbol \cite{nos}.  After torsion has been
eliminated, there is no ambiguity for our analysis which only has
extremals at its disposal. That interaction with extremals,
however, crucially depends on the choice of the 'physical' metric
to be used: the one computed from the $e_\mu^a$ of (1), or any
$\hat g_{\alpha\beta}$ which is a result of different field
transformations involving the dilaton field? Clearly the
torsionless dilaton theory (\ref{Gl:22}) has the same metric as
(\ref{Gl:1}), e.g.
the global analysis of \cite{kat93} applies directly and the
different types of solutions are exhausted by those studied in
\cite{kat93}.

However, from the point of view of a 'true' dilaton theory, one
could argue that with a redefined metric as in (\ref{Gl:25}),
 $\hat g_{\alpha\beta} =
e^{2\phi}g_{\alpha\beta} = 2g_{\alpha\beta} / X$ has some physical
justification as well.  In fact, for Witten's black hole
$g_{\alpha\beta}$ is flat and the interesting (black hole)
singularity structure  just results from the factor $2/X$.  Now, in
the original $R^2 + T^2$--theory \cite{kat93} there are solutions
(G3) resembling e.g.  the black hole but not completely: Their
singularity resides at light--like lines and they are not
asymptotically flat in the Schwarzschild sense.  Thus the factor
$2/X$ may well yield improvements on that situation. \\

Here we shall analyse a generalized dilaton gravity
\begin{equation}
\label{Gl:28}
{\cal L} = \sqrt{-\hat g} e^{-2\phi} \left[\hat R +
4(1-2\alpha e^{-2\phi}) (\nabla \phi)^2 + 2\beta e^{-4\phi} + 4\lambda^2
\right]
\end{equation}
which is obtained from (\ref{Gl:17}) by taking
\begin{equation} \label{Gl:29}
X = 2e^{-2\phi}~~~~
g_{\mu \nu} = {\hat g}_{\mu \nu} e^{-2\phi}~~~~
\Lambda = -4\lambda^2
\end{equation}
and omitting an overall minus sign.
We need to consider only the cases for $\beta$ = positive, negative
or $0$. The absolute value of a nonvanishing $\beta$ may always be
absorbed by rescaling $X$ and $\omega$ to $X \rightarrow
\sqrt{\vert\beta\vert}X$
and $\omega \rightarrow \frac{\omega}{\sqrt{\vert\beta\vert}}$.
Let us start with a positive value for $\beta$ e.g. $+2$.
All global solutions with nonconstant curvature
are most easily obtained by the known general
solution \cite{sch93} of
the equations of motion of (\ref{Gl:17}) for the zweibein $e^a$
$(X^+ \ne 0)$  in an arbitrary gauge
\begin{equation}
\label{Gl:30}
e^+ = X^ + e^{\alpha X} df
\end{equation}
\begin{equation}
\label{Gl:30b}
e^- = \frac{dX}{X^+} + X^- e^{\alpha X} df
\end{equation}
where $f$, $X$ and $X^+$ are arbitrary functions except for the requirement
that $df$ and
$dX$ define a basis for one forms.
The line
element from (\ref{Gl:30}) and (\ref{Gl:30b}) reads
\begin{equation}
\label{Gl:31}
(ds)^2 = 2 e^+ e^- = 2e^{\alpha X} df\otimes \left( dX+X^+X^-e^
{\alpha X} df \right).
\end{equation}
As shown in \cite{kat86} the Lagrangian in all such models gives rise to an
absolutely
conserved quantity
\begin{eqnarray}
\label{Gl:32}
C &=& X^+X^-e^{\alpha X} + w(X)
\end{eqnarray}
\begin{eqnarray}
\label{Gl:32b}
w(X) &=& \int_{X_0}^{X} V(y) e^{\alpha y} dy,
\end{eqnarray}
where the lower limit $X_0 = const$ clearly
has to be determined appropriately
so that (inside a certain patch) the integral exists for a certain
range of X.
$C$ can be used to eliminate $X^+ X^-$ in (\ref{Gl:31}).
Using  (\ref{Gl:29}) and defining coordinates $v = -4f$, $u =
\phi$  yields
the line element of the generalized dilaton Lagrangian
(\ref{Gl:28})
\begin{equation}
\label{Gl:33}
(d \hat{s})^2=e^{2u} (ds)^2 = g(u) \left( 2dvdu+ l(u) dv^2 \right)
\end{equation}
with
\begin{equation}
\label{Gl:34}
l(u)=\frac{e^{2u}}{8} \left(
C- g(u) \left( \frac{4e^{-4u}}{\alpha} -
\frac{4e^{-2u}}{\alpha^2}+ C_0 \right) \right)
\end{equation}
\begin{equation}
g(u)= e^{2\alpha e^{-2u}},
\end{equation}

\begin{equation}
\label{Gl:38}
C_0 =  \frac{2}{\alpha^3}
+ \frac{4\lambda^2}{\alpha}
\end{equation}
which automatically implies the convention for the  constant of
integration in (\ref{Gl:32b})
to be used in the following.

Note that transformation (\ref{Gl:21}) is defined only for positive $X$ while
the
original model contains arbitrary values of $X$. To cover the negative case one
would
have to replace (\ref{Gl:21}) by
\begin{equation}
\label{Gl:21b}
\frac{X}{2} = -e^{-2\phi},~~~~X<0.
\end{equation}
Since the metric (\ref{Gl:33}) is invariant under
\begin{equation}
\label{Gl:invar}
e^{-2u} \rightarrow -e^{-2u},~~~~\alpha \rightarrow -\alpha,~~~~C \rightarrow
-C
\end{equation}
one may consider the line element (\ref{Gl:33}) for $\alpha>0$ and $\alpha<0$
to cover
all patches of the original $R^2+T^2$-theory where $X$ is positive and
negative.

In the metric  of the form
\begin{equation}
\label{Gl:39}
{\hat g}_{\mu \nu}=g(u)\left( \begin{array}{rr}
                    0   &   1     \\
                    1   &   l(u)
\end{array} \right).
\end{equation}
the Killing direction is $\partial/ \partial v$ and the norm
of the Killing vector
$$
k^{\mu}=\left( \begin{array}{r}
                       0 \\
                       1
\end{array} \right)
$$
becomes $g(u)l(u)$. By a redefinition of the variable $u$ it would
be easy to reduce (\ref{Gl:39}) to the standard light cone gauge with $ g = 1$,
used e.g. in the second ref. \cite{kat93}. However
because of $g(u)>0$ (except at $u \rightarrow \infty$)
the horizons are determined by the zeros of $l(u)$ only.
Therefore, we  have found it technically  easier to retain
(\ref{Gl:39}). \\

The conformal gauge $(d \hat s)^2 = F(u)\, d\tilde {u}'\,
d\tilde {v}'$ in (\ref{Gl:39}) is obtained by
'straightening' the null extremals
\begin{equation}
\label{Gl:40} v=const.
\end{equation}
\begin{equation}
\label{Gl:41} \frac{dv}{du}=- \frac2l \mbox{ for all $u$ with $l(u) \neq 0$}
\end{equation}
\begin{equation}
\label{Gl:42} u=u_0 = const \mbox{ for $l(u_0)=0$}
\end{equation}
by means of a diffeomorphism
\begin{eqnarray}
\label{Gl:43}
\tilde {u}' = v+f(u),~~~~\tilde {v}'= v \\
f(u) \equiv \int^u \frac{2 dy}{l(y)}.
\end{eqnarray}
A subsequent one  $\tilde {v}' \rightarrow \tan \tilde
{v}'$
and another  appropriately chosen one
for $\tilde {u}'$ produce  the Penrose diagram. It is valid  for a
certain patch where (44) is well defined.
Clearly the shape of those diagrams
depends crucially on the (number and kind of) zeros and on the asymptotic
behavior of $l(u)$.

\section{Classification of Global Solutions}
The analysis of all possible cases as described
by the ranges of parameters $\alpha, \beta, C$ and $\lambda^2$ is
straightforward but tedious.
We shall first give the classification of all global solutions.
Comments on the construction of Penrose diagrams will be presented in the
next Section.
Apart from $C_0$, defined in (38),
also
\begin{equation}\label{Gl:45}
C_1 = \frac{2}{\alpha^2}\; e^{2\alpha\sqrt{-\lambda^2}}\;
(\frac{1}{\alpha} - 2\, \sqrt{-\lambda^2})
\end{equation}
plays a role for $\alpha \neq 0$ and $\lambda^2 < 0$, discriminating the
possible cases with two zeros, with one double--zero and without
zero in $l$, i.e. the presence of two nondegenerate or one
degenerate killing--horizon. The qualitatively distinct cases for
$\alpha > 0$ and $\alpha < 0$ are listed in (\ref{Gl:46}) and
(\ref{Gl:47}):

\underline{{$\boldmath{\alpha > 0,~~~\beta > 0:}$}}
\begin{equation} \label{Gl:46}
\begin{array}{l}
D1^+:~~~C> C_0   \\
D2^+:~~~C= C_0,
{}~~~~\lambda^2 < 0 \\
D3^+:~~~C= C_0,
{}~~~~\lambda^2 \geq 0 \\
D4^+:~~~C< C_0,
{}~~~~C> C_1,
{}~~~~\lambda^2 <0 \\
D5^+:~~~C< C_0,
{}~~~~C= C_1,
{}~~~~\lambda^2 <0  \\
D6^+:~~~C< C_0,
{}~~~~C< C_1,  \\
\end{array}
\end{equation}

\underline{{$\boldmath{\alpha < 0,~~~\beta > 0:}$}}
\begin{equation} \label{Gl:47}
\begin{array}{l}
D1^-:~~~C>C_0 ,~~~C \geq 0 \\
D2^-:~~~C>C_0 ,~~~C < 0 \\
D3^-:~~~C=C_0 ,~~~C \geq 0 \\
D3_r^-:~~~C=C_0 ,~~~C <0 ,~~~ \lambda^2 \geq 0 \\
D4^-:~~~C=C_0 ,~~~C <0 ,~~~ \lambda^2 < 0 \\
D1_r^-:~~~C<C_0 ,~~~C< C_1 \\
D5^-:~~~C<C_0 ,~~~C= C_1 \\
D6^-:~~~C<C_0 ,~~~C> C_1, ~~~C<0 \\
D2_r^-:~~~C<C_0 ,~~~C> C_1, ~~~C \geq 0\\
\end{array}
\end{equation}
with the index $r$ indicating a rotation by $90^0$.
The classification $Dn^\pm$ denotes the Penrose diagrams for
generalized dilaton gravity.
Figs.(1) and (2) show their range of parameters
and in Figs.(3) and (4) the corresponding
Penrose diagrams are depicted. These cases for nonzero $\alpha$ are locally
equivalent
to $R^2+T^2$-gravity defined by the actions (\ref{Gl:1}) and (\ref{Gl:17}).

The limit $\alpha =0$ -- corresponding to ordinary dilaton gravity with an
additional
potential $\beta e^{4 \phi}$ -- is defined only for the action in first order
form (\ref{Gl:17}). It is equivalent to
(higher derivative) $R^2$-gravity without torsion,
whose global solutions are analyzed in the second reference of  \cite{kat93}.
To obtain the classification of the global solutions for
the locally equivalent dilaton theory we use (\ref{Gl:31}) and (\ref{Gl:32})
with $\alpha =0$ and again $\beta =2$ to get the line element
\begin{equation}
\label{Gl:48}
(d \hat s)^2 = 2dudv + e^{2u} \left( \frac{C}{8}- \lambda^2 e^{-2u} -
\frac{e^{-6u}}{3} \right) dv^2.
\end{equation}
In terms of
\begin{equation}
\label{Gl:49}
C_1^0 \quad = \quad -\frac{16}{3}\; \vert
\lambda^2\vert^{3/2}
\end{equation}
qualitatively this yields exactly the same cases
as for $\alpha >0$
for the following conditions on the parameters:

\underline{{$\boldmath{\alpha = 0,~~~\beta > 0:}$}}
\begin{equation}\label{Gl:50}
\begin{array}{l}
D1^+:~~~C> 0 \\
D2^+:~~~C= 0,~~~~\lambda^2<0 \\
D3^+:~~~C= 0,~~~~\lambda^2 \geq 0 \\
D4^+:~~~C< 0,~~~~C>C_1^0\\
D5^+:~~~C< 0,~~~~C=C_1^0\\
D6^+:~~~C< 0,~~~~C<C_1^0\\
\end{array}
\end{equation}

A sign change of $\beta$ to $-\beta$ together with a sign change of $C$ and
$\lambda^2$, followed by $dv \rightarrow -dv$ produces the same global
solutions
except that timelike and spacelike directions are reversed
$(ds^2 \rightarrow -ds^2)$. Since all possible signs for $C$ and $\lambda^2$
were considered above we obtain the solutions for negative $\beta$'s simply by
rotating the diagrams by $90^0$ and keeping the sign change of $C$ and
$\lambda^2$ in mind.

Next we consider
the teleparallel limit of (\ref{Gl:17}) corresponding to $\beta =0$.
In this limit the curvature is equal to zero and one has $T^2$--gravity
\cite{obuk}.
We still obtain the line element of (\ref{Gl:33}) with a
simplified
\begin{equation}
\label{Gl:simple}
l=\frac{e^{2u}}{8} \left( C-e^{2\alpha e^{-2u}}
\frac{4\lambda^2}{\alpha}\right).
\end{equation}

The Penrose diagrams for the corresponding global solutions have already
been obtained in the
above cases. Here they are related to the following ranges of parameters
$(D=C-\frac{4\lambda^2}{\alpha})$:
\\
\underline{{$\boldmath{\alpha \neq 0,~~~\beta = 0:}$}}
\begin{equation}\label{Gl:50b}
\begin{array}{l}
D6_r^+:~~~D> 0,~~~~\lambda^2<0 \\
D1^-:~~~D> 0,~~~~\lambda^2=0 \\
D1^+:~~~D> 0,~~~~\lambda^2>0 \\
D1_r^+:~~~D< 0,~~~~\lambda^2<0 \\
D1_r^-:~~~D< 0,~~~~\lambda^2=0 \\
D6^+:~~~D< 0,~~~~\lambda^2>0 \\
D3^+:~~~D= 0,~~~~\lambda^2>0\\
D3_r^+:~~~D= 0,~~~~\lambda^2<0\\
D1^-:~~~D> 0,~~~~C\geq 0 \\
D2^-:~~~D> 0,~~~~C<0 \\
D2_r^-:~~~D< 0,~~~~C> 0 \\
D1_r^-:~~~D< 0,~~~~C\leq 0 \\
D3^-:~~~D= 0,~~~~\lambda^2<0 \\
D3_r^-:~~~D= 0,~~~~\lambda^2>0 \\
\end{array}
\end{equation}
\\
The case $\lambda=0, C=0$ yields just
the conformally flat case.

The last case $\alpha = \beta = 0$ of (\ref{Gl:29}) describes ordinary dilaton
gravity.
Its general solution is
\begin{equation}
\label{last}
(d \hat s)^2 = 2dudv + e^{2u} \left(C e^{2u}- \lambda^2  \right) dv^2
\end{equation}
and contains three types of global solutions:\\
\underline{{$\boldmath{\alpha = \beta = 0:}$}}
\begin{equation}\label{Gl:null}
\begin{array}{l}
D2^+:~~~C> 0,~~~~\lambda^2>0 \\
D2_r^+:~~~C< 0,~~~~\lambda^2<0\\
D3^+:~~~C> 0,~~~~\lambda^2 \leq 0 \\
D3_r^+:~~~C< 0,~~~~\lambda^2 \geq 0 \\
\end{array}
\end{equation}
and flat space-time for $C=0$.

\section{Construction of Penrose Diagrams}
For each set of the parameters as summarized in (\ref{Gl:46}),
(\ref{Gl:47}), (\ref{Gl:50}), (\ref{Gl:50b}) and (\ref{Gl:null})
the global solution is obtained
by gluing (whereever necessary) solutions of the line element (\ref{Gl:33}),
which
as a rule
defines only a local solution.
Another local solution  in conformal coordinates is obtained by
interchanging the role of the null--directions. The transformation
\\
\parbox{12cm}
{\begin{eqnarray*}
\tilde{u}'' &=& u \\
\tilde{v}'' &=& -f(u) - w
\end{eqnarray*}}
\hfill
\parbox{1.0cm}{\begin{eqnarray}\label{Gl:52}\end{eqnarray}}\\
with $f(u)$ from (44) may be easily verified to do this job. The
situation may be visualised best in a specific example, say
$D2^+$ \cite{not} as shown in Fig. 5. Eqs. (\ref{Gl:52}) essentially
correspond to
a reflection of the diagram with respect to an axis orthogonal to
the lines $u = const$, i.e. transversal to the Killing directions $u
= const$.  In Fig.5 this leads from a) to b), resp. c). Further
solutions are obtained by simple reflections in the ($\tilde u,
\tilde v$)--coordinates. Now those patches may be glued together
along certain parts of their boundaries by identifying either the
triangle or the square. For that purpose the completeness of all
extremals at these edges must be analysed by checking whether there
is at least one extremal reaching a certain boundary (or corner of
a boundary) at a finite
value of the affine parameter.  By definition, a point not to
be reached by any extremal is complete. Using the conservation law for a
metric of type (\ref{Gl:39}) from its Killing direction

\begin{equation}
\label{Gl:53}
g\; \frac{du}{d\tau} + gl\frac{dv}{d\tau} = \sqrt{A} = const\; ,
\end{equation}
and identifying the affine parameter $d\tau$ with the $d \hat s$ in
(\ref{Gl:33}), these extremals are found to obey

\begin{equation}
\label{Gl:54}
\frac{dv}{du} = -\frac1l \left[ 1 \mp ( 1 - l
g/A)^{-1/2}\right]
\end{equation}
by simply solving a quadratic equation. In addition,  from
(\ref{Gl:53}) and (\ref{Gl:54}) the affine parameter is determined by
\begin{equation}
\label{Gl:55}
\hat s(u)= \frac{1} {\sqrt{\vert A \vert}}\int^{u} dy\;
 g \left( 1 - \frac{lg}{A} \right) ^{-1/2} ~~~ .
\end{equation}
For $A > 0$, resp. $ A < 0$ the parameter $s$ is a timelike, resp.
spacelike quantity.
Extremals along the null--directions (\ref{Gl:40}), (\ref{Gl:41})
are contained
as the special case $A \to \infty$ in (\ref{Gl:54}), with (\ref{Gl:55})
replaced by
similar relation without the square--root terms.
In addition to (\ref{Gl:53}), (\ref{Gl:54}) and (\ref{Gl:55})
the equation for extremals is satisfied by the horizons (\ref{Gl:42})
and by degenerate extremals (see (\ref{Gl:degener}) below) for which
$lg=A$ and therefore (\ref{Gl:55}) does not hold.
Clearly the additional
condition $l'\,(u_0) = 0$ identifies double zeros (degenerate
Killing horizons) as e.g. in $D5^-$. Incompleteness of the lines
$u_0 = const$ for single zeros at one end is established easily. A line at a
double zero as in $D5^-$ instead  has incomplete endpoints. From
considering the extremals for finite $A \neq 0$, in all cases the
edges with $u = \pm \infty$ are found to be incomplete, {\sl except}
for $C = C_0$ (cf. e.g. our example $D2^+$ or $C = 0$ in $D2^0$) where for
 $u \to + \infty$ the boundary is complete.
We remark that the completeness of null and non-null extremals differs on some
boundaries.
For example at $u \rightarrow -\infty,~~\alpha>0$ null extremals are complete
while
non-null extremals are incomplete. Contrary to this situation the completeness
of all types
of extremals is the same in $R^2+T^2$-theory. It can be shown that this
difference is due
to the conformal rescaling of the metric (\ref{Gl:23}).

Except for the cases $D2^+$, $D3^{\pm}$,
$D3_r^-$, $D4^-$ (where $R \to 0$ for $u \to +\infty$)
the scalar
curvature diverges at
$u \rightarrow \pm \infty$ which can be seen directly evaluating
\begin{equation}
\label{Gl:51}
R=\frac{1}{g^2} \left( \frac{g,_{u}^2}{g} l -g,_{uu} l -
g,_u l,_u -g l,_{uu} \right)
\end{equation}
and using (\ref{Gl:34}).
Thus, all singular boundaries are incomplete while the boundaries with zero
curvature
are always complete. This means that generalized dilaton gravity includes
assymptotically flat solutions, something not encountered in the original
$R^2+T^2$-theory.

Special consideration require the corners of a boundary
formed by lines with both $u = +\infty$ or
both $u = -\infty$.
For example, in Fig.5 the lower right one
can only be reached by extremals
(\ref{Gl:53}) with
(\ref{Gl:55}) when $A = 0$. Actually for $D2^+$  that point is  complete ---
just as the adjoining lines $u = +\infty$.
In some diagrams such corners are formed by lines $u=+\infty$
and $u=-\infty$ both singular in curvature. These corners are essential
singularities and can only be reached by degenerate extremals, which are
parallel to the $v$-axis and go through those points where
\begin{equation}
\label{Gl:degener}
\left. \frac{\partial (lg)}{\partial u} \right|_{u=u_0} =0
\end{equation}
is satisfied. The existence and number of such extremals depends on the values
of the parameters. They are always complete and so are the corresponding
corners.
If (\ref{Gl:degener}) has no solution then the corner is not reached by any
extremal and thus is complete by definition.
In Figs.\ 3 and 4 these
points are indicated by full dots. With
these tools, patches as in Fig.\ 5 may now be glued together. For
$D2^+$ this leads to the well--known shape of the 'classical' black
hole in the corresponding global solution drawn in Fig.\ 3.
A full line in this diagram ---
as in the others --- represents an (incomplete) singularity of the
curvature at $u = \pm \infty$, a thin line at the external boundary
denotes a  complete  asymptotically flat space-time which may occur at
$u = +\infty$ only. Internal lines indicate $u = const$, with a broken
line for a Killing horizon. Arrows indicate directions into which one
should imagine periodic continuation.\\

\section{Comparison of the Models}
It seems instructive to compare our present global structure to the
one studied for other theories.
The basic observation is that each global solution
of the original $R^2+T^2$-theory naturally
splits into a set of global solutions in
generalized dilaton gravity.
As a generic example we show
in Fig. 6 (a),(b) and (c) the three different ways of
splitting of the diagram $G3$ (in the notation of \cite{kat93})
where the seperation is determined solely by
the value of the constants. The Penrose diagram $G3$
covers all values of $X$. Since the dilaton field is introduced seperately for
positive and negative $X$ the Penrose diagram $G3$ is split by the lines $X=0$,
where the conformal transformation in (\ref{Gl:29}) becomes singular.
The resulting diagrams
directly turn into
the global solutions of generalized dilaton gravity.
It should be noted that
in the $R^2+T^2$--theory all non-null boundaries can be straightened by a
suitable
choice of coordinates, while in generalized dilaton gravity some boundaries
cannot be straightened without affecting other boundaries. For example in the
square
diagram $D1^+$ or in the "eye"-diagram $D6^+$ a redefinition of the coordinates
cannot straigthen all boundaries simultaneously, whereas this is possible in
the diagram $D2^-$. This phenomenon has been observed before
(cf. the second reference of \cite{kat93}).

It is also important that the conformal transformation of the metric
${\hat g}_{\alpha \beta} = 2 g_{\alpha \beta}/X$ changes the character
of the boundaries.
The boundary $X=0 ~~(u \rightarrow +\infty)$ becomes singular
$\tilde{R} (\hat g) \rightarrow \infty$, and remains incomplete for $C \neq
C_0$,
while for $C = C_0$ it becomes complete and assymptotically flat
$\tilde{R} (\hat g) \rightarrow 0$.
The boundaries $X \rightarrow \pm \infty$ remain singular but always
become incomplete.

In Fig.6(d) the "black hole" solution $D2^+$ emerges.
It illustrates how the
diagram $G11$ that had to be continued in the plane splits into the diagrams
$D2^+$ and
$D4^-$ which cannot be continued.
The decomposition in the case of ordinary dilaton gravity is shown in Fig.7.
Model (\ref{Gl:17}) for $\alpha = \beta =0$ has a unique global
solution which is flat Minkowskian space-time represented by
a rhombic Penrose diagram. Performing the conformal transformation
yields a singular metric and curvature at the lines of vanishing $X$
thereby producing three global solutions of the dilaton model.
$D2^+$ represents the famous black hole solution \cite{wit91}
whose singularity is reached by time-like extremals for a
finite value of the affine parameter.
The completeness of the other extremals is the same as in the above cases.
The original $R^2+T^2$--theory
contains one solution resembling the 'real' Schwarzschild black hole
only in a very approximate sense.
Its solution $G3$ exhibits an (incomplete) singularity, but into
null--directions, the 'asymptotically flat' direction is replaced
by a singularity of the curvature, albeit at an infinite
distance (complete case).
In the present dilaton theory
precisely the example $D2^+$,
whose derivation from $G11$ was
discussed more explicitly above, is Schwarzschild--like.
Other solutions with similar properties, but
more complicated singularity structure are $D3^-$
(naked singularities) and $D4^-$. On the
other hand, the 'eye' diagram $D6^+$ appears here, as well as the
square diagrams $D1^+$ of $R^2$--gravity \cite{kat93}.  $D4^-$
represents an interesting variety of a manifold where the ordinary
black hole is replaced by a 'light'--like singularity.

\section{Summary and Outlook}

By showing explicitly the local equivalence of certain (torsionless)
dilaton theories with a 2d theory quadratic in curvature and
torsion, a certain class of such generalizations of the original
Witten--model has now been found to acquire a better geometric foundation.
At the same time,
however, we observe that local equivalence does not guarantee the equivalence
of global solutions. Indeed, we have shown
in two generic examples how
one global solution
of 2d gravity with torsion splits into a set of global solutions of
generalized dilaton gravity.
This seperation occurs along lines $X=0$
where the conformal transformation of the metric
${\hat g}_{\alpha \beta} = 2 g_{\alpha \beta}/X$
becomes singular \cite{mignemi}.

We find that in the equivalent dilaton theories one of the main
advantages of the original "minimal" dilaton theory
\cite{wit91}, the presence of
an asymptotically flat black
hole solution, resembling the 4d--case, is
retained. Since the
complete classical solution is known in all cases, the range of
models thus has been extended considerably for which in a next step quantum
effects can be studied, after interactions with matter have been
added. Of course, we encounter here once more the problem, familiar
from Brans--Dicke--Jordan type theories \cite{bran}: Our analysis
implicitly assumes that our 'testparticle' obeying the equations for
the extremals in Section 6 for the 'equivalent theory' is coupled
to a metric of the Jordan--version of the theory, i.e., to the
redefined metric $\hat g$ and not to the metric
derived from the original $R^2+T^2$--theory. It must be admitted
that therefore the argument of geometric interpretability is
somewhat weakened. Nevertheless, we are convinced that the class of
such models introduced here may serve as a field for promising
further studies.

\section*{Acknowledgement}

We are grateful for discussions with T. Kl\"osch and M. Nikbakht.
This work has been supported by Fonds zur F\"orderung der
wissenschaftlichen For\-schung (FWF) Project No.\ 10221--PHY.  One
of the authors (M.K.) thanks The International Science Foundation,
Grant NFR000, and the Russian Fund of Fundamental Investigations,
Grant RFFI--93--011--140, for financial support.

\newpage


\begin{center}
\leavevmode
\epsfysize=7cm
\epsfbox{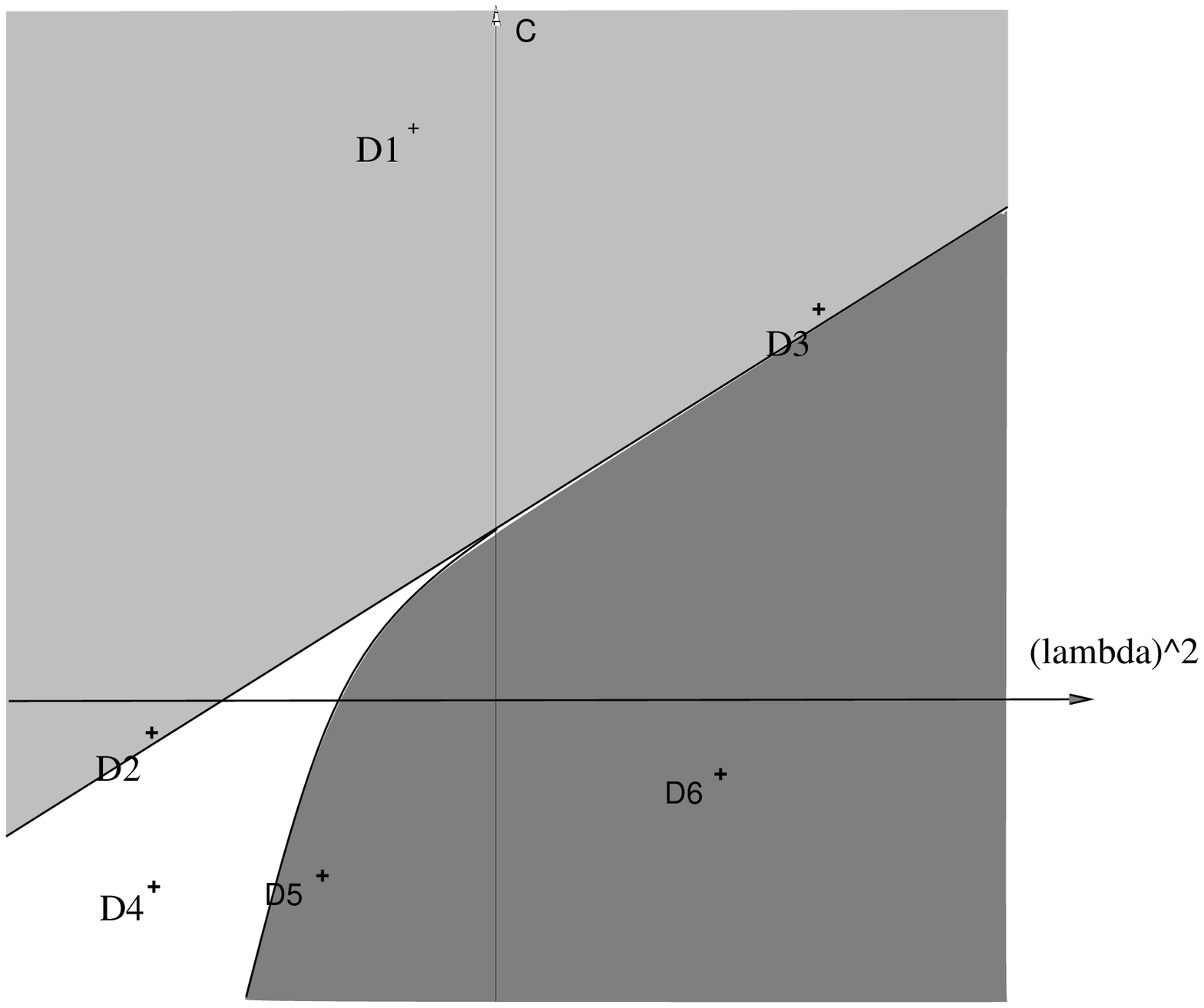}
\\
\centerline{Fig.\refstepcounter{figure}$\;$ \thefigure
\smallskip {}~~~Range of solutions for $\alpha > 0$}
\label{Fig1}
\end{center}
\vspace{.25in}

\begin{center}
\leavevmode
\epsfysize=7cm
\epsfbox{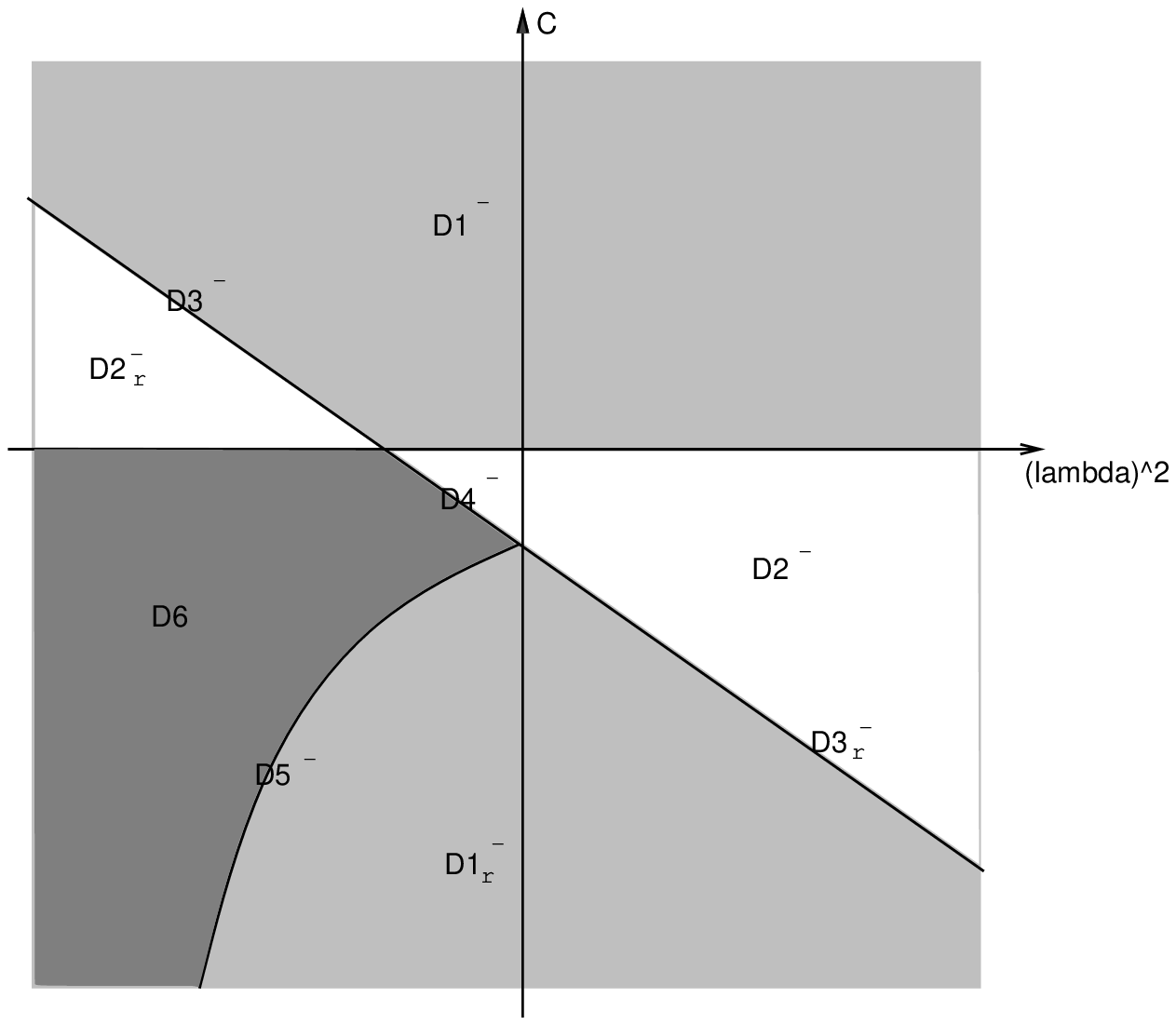}
\\
\centerline{Fig.\refstepcounter{figure}$\;$ \thefigure
\smallskip {}~~~Range of solutions for $\alpha < 0$}
\label{Fig2}
\end{center}
\vspace{.25in}

\begin{center}
\leavevmode
\epsfysize=8cm
\epsfbox{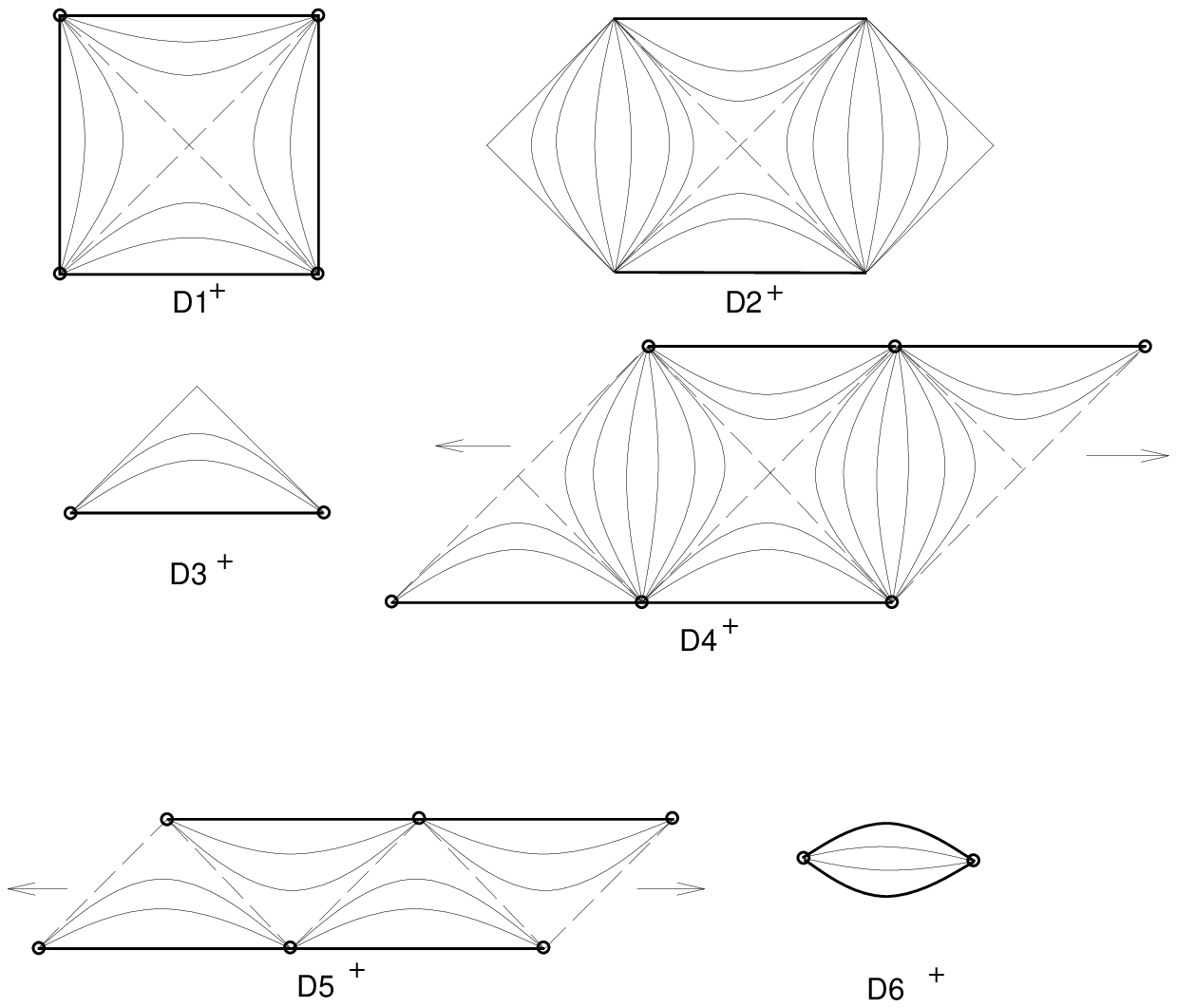}
\\
Fig.\refstepcounter{figure}$\;$\thefigure
{}~~Penrose diagrams for $\alpha >0$. Thick lines indicate an incomplete
singular
boundary $(R \rightarrow \infty )$. Thin lines show (complete) boundaries with
finite
or vanishing curvature,  inside the diagrams  they correspond to lines of
constant curvature. Dots  represent complete corners and Killing
horizons are drawn as dashed lines. \label{Fig3}
\end{center}
\vspace{.25in}

\begin{center}
\leavevmode

\epsfysize=8cm
\epsfbox{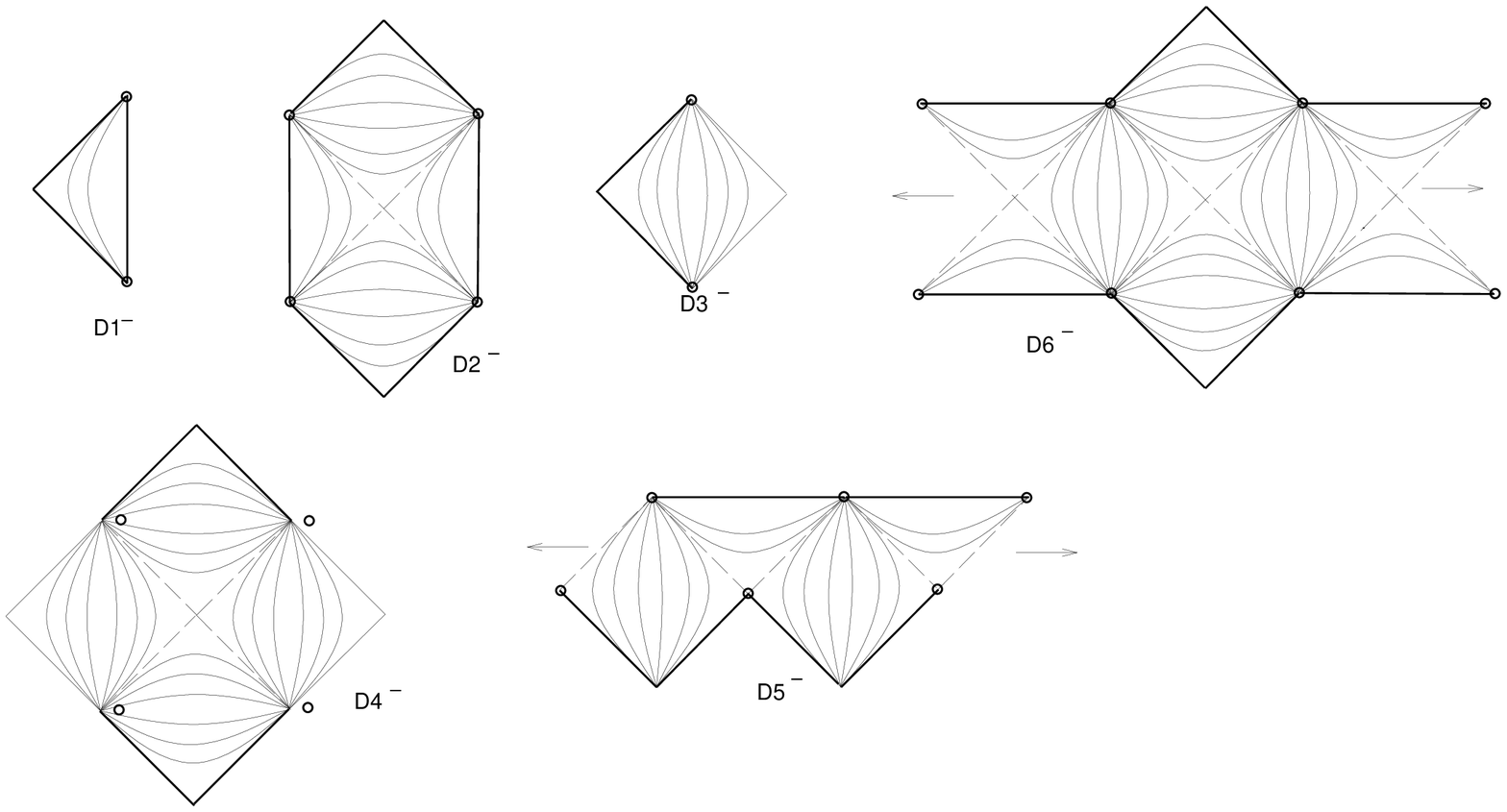}

Fig.\refstepcounter{figure}$\;$ \thefigure
{}~~~Penrose diagrams for $\alpha<0$. Conventions as above.
\label{Fig4}
\end{center}

\begin{center}
\leavevmode
\epsfxsize=16cm
\epsfbox{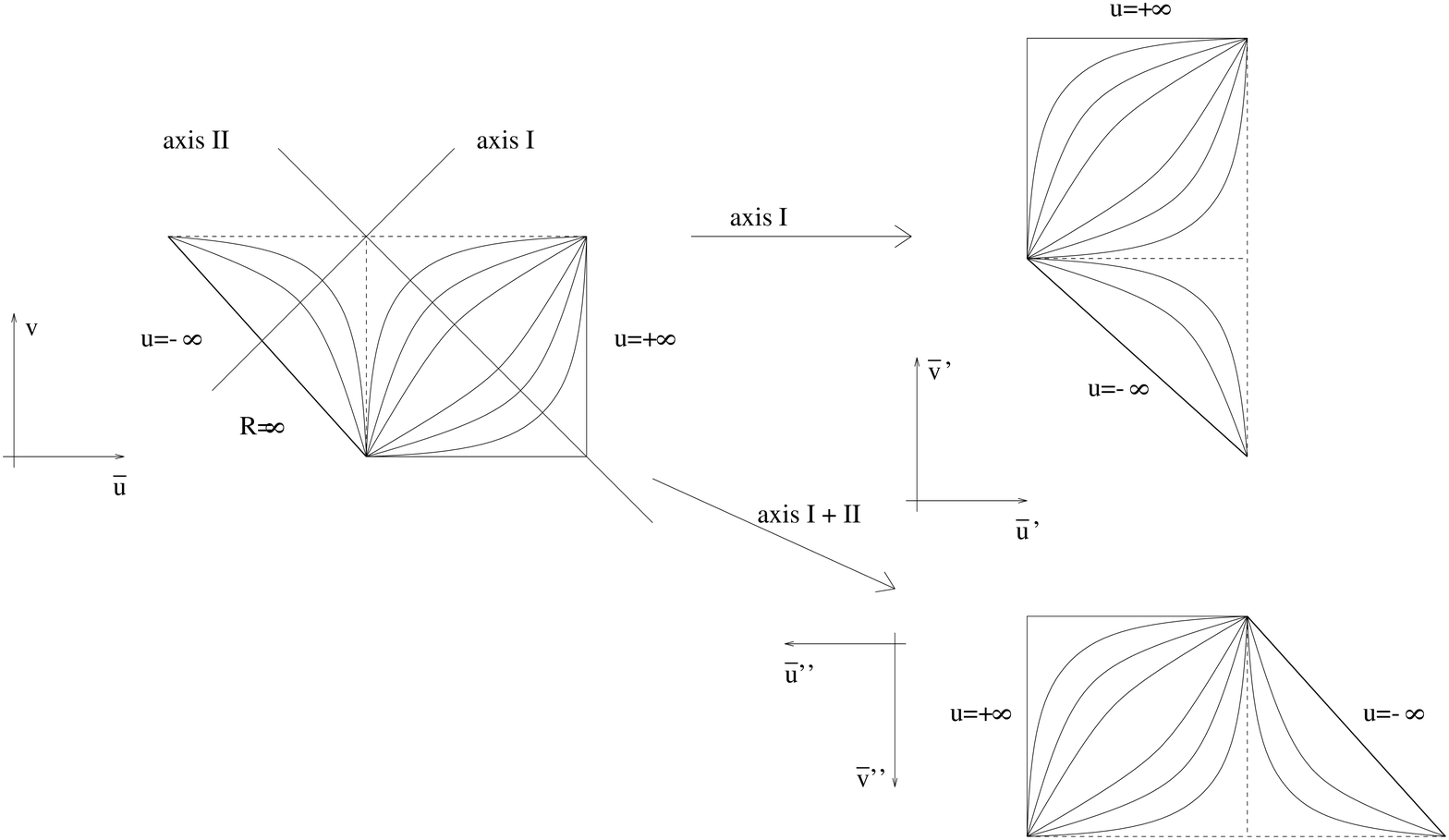}
\\
Fig.\refstepcounter{figure}$\;$\thefigure
{}~~~Different types of solutions for $D2^+$
\label{Fig5}
\end{center}
\vspace{.25in}

\begin{center}
\leavevmode
\epsfxsize=12cm
\epsfbox{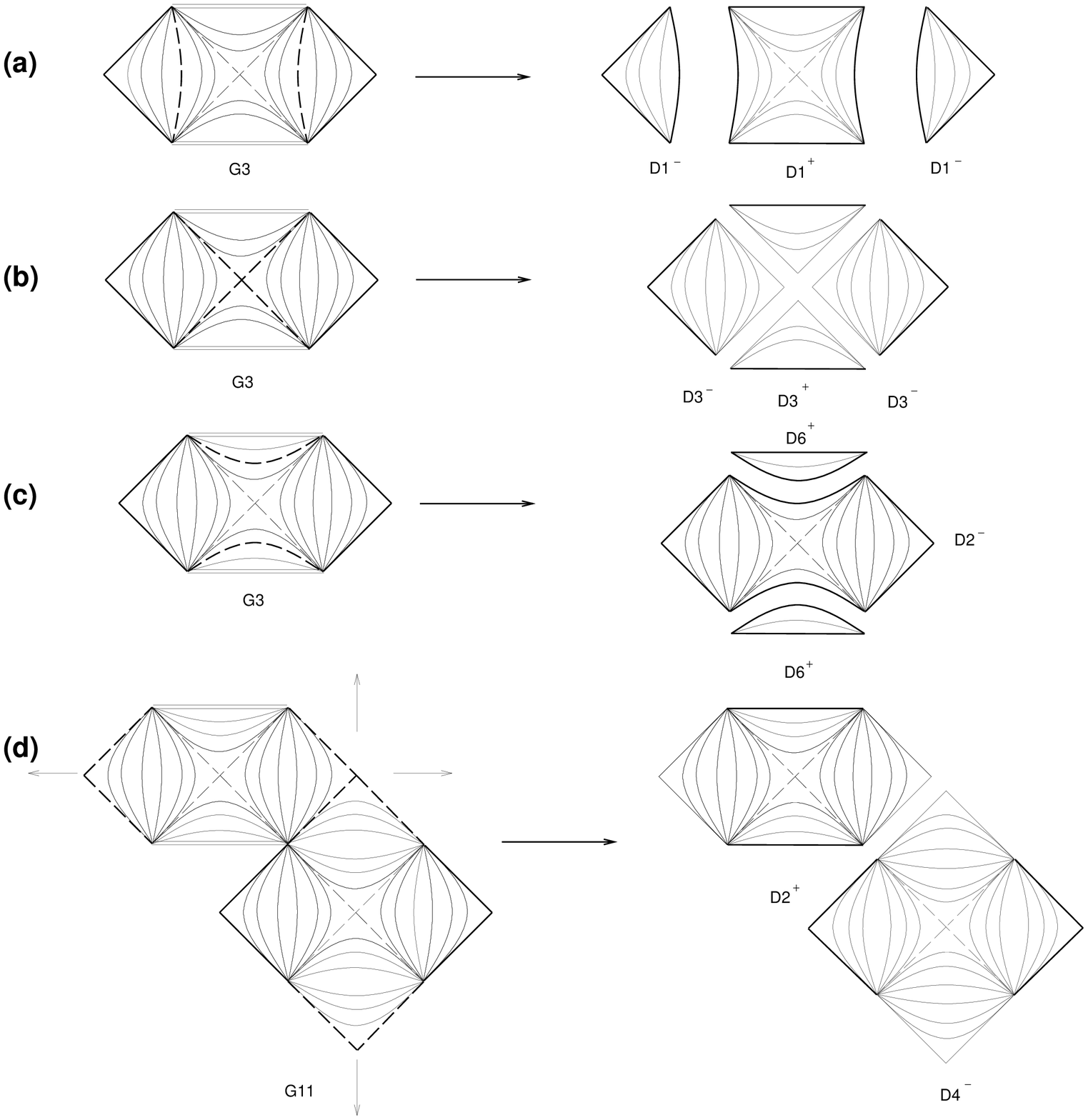}
\\
Fig.\refstepcounter{figure}$\;$\thefigure
{}~~~Examples of the splitting of the $R^2 + T^2$ diagrams into the diagrams of
generalized dilaton theory. The double lines indicate complete
singular boundaries and thick dashed lines correspond to $X=0$.
(a), (b) and (c)
show different splittings of the same diagram G3 for $\lambda^2 >0$ which
depend on the parameters as follows:
(a) $\alpha C_0 < \alpha C$, (b) $C=C_0$, (c) $0<\alpha C < \alpha C_0$.
(d) occurs for $\lambda^2<0,~~\alpha C=\alpha C_0>0$ and demonstrates the
appearance
of the black hole solution $D2^+$ in the generalized dilaton theory.
\label{Fig6}
\end{center}
\vspace{2.25in}

\begin{center}
\leavevmode
\epsfxsize=12cm
\epsfbox{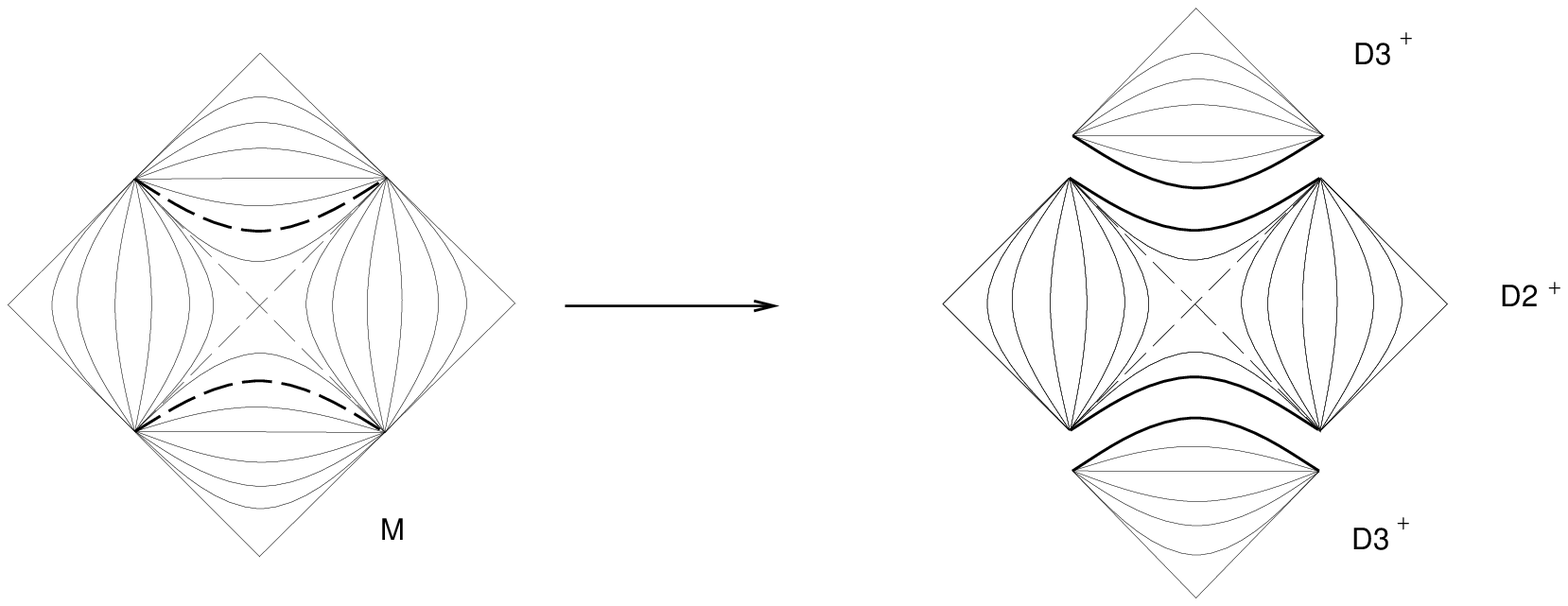}
\\
Fig.\refstepcounter{figure}$\;$\thefigure
{}~~~Splitting of Minkowskian space-time $M$ into the black hole of ordinary
dilaton gravity. The thin lines in the left diagram are the lines of $X=const$
and
especially the thick dashed line denotes $X=0$ along which the splitting
occurs.
\label{Fig7}
\end{center}
\vspace{.25in}


\begin{thebibliography}{99}
\bibitem[9]{nos}No simple Lagrangian interpretation exists for 'geodesics'
		(autoparallels), defined with respect to affine connections including also
		the cotorsion contribution, if torsion is nonvanishing.
\bibitem[13]{XY}Actions of this type have been widely discussed before (cf.
		\cite{ban91}).
\end{thebibliography}
\end{document}